\documentclass[english,aps,prstper,reprint,showpacs,titlepage,longbibliography]{revtex4-2}   

\usepackage[T1]{fontenc}	% should generally be included for better accented-word behavior
\usepackage{geometry}		% for controlling page margins
\usepackage{graphicx}
\usepackage[above,below]{placeins}	% allows use of \FloatBarrier command to force section barriers
\usepackage{times}

\usepackage{hyperref}  
\hypersetup{colorlinks=true,urlcolor=blue,citecolor=blue,linkcolor=blue}   
\usepackage{enumitem}          % package for handling list formatting
\setlist{nosep}                 % Tightest spacing for lists. `noitemsep` is more relaxed

\begin{document}

\begin{titlepage}

\title{P-A.I.R.: A Structured AI-Replication Framework for Active Learning in Introductory Physics}
 \author{Bilas Paul}
\affiliation{Department of Physics, SUNY Farmingdale State College, Farmingdale, NY, 11735}

\begin{abstract}
\label{abstract}
\noindent 
The growing use of generative AI tools among students raises an important pedagogical question: how can AI be structured to promote active learning rather than passive answer-seeking? This study introduces the Physics AI-Replication (P-A.I.R.) framework, in which students identify challenging problems, use AI to explain underlying concepts and solutions, generate similar problems, and practice independently before reviewing answers. Survey data from 39 undergraduate students in algebra-based physics indicate that nearly all participants reported improved conceptual understanding following engagement with P-A.I.R. across the semester, and self-confidence scores were consistently above the scale midpoint ($\mu = 7.03$; range 5--9). A strong association between conceptual clarity and replication helpfulness ($r_s = 0.582$, $p < 0.001$) supports the framework’s design. Qualitative findings highlight conceptual clarification and structured problem-solving. These results suggest that P-A.I.R. offers a practical approach for integrating AI into physics learning.
\clearpage
\end{abstract}

\maketitle
\end{titlepage}

%========================================================================================================================================================%
\section{Introduction }
\label{intro}
%========================================================================================================================================================%
The rapid emergence of generative artificial intelligence (AI) tools has begun to reshape how students engage with learning across disciplines, including physics. Systems such as large language models can provide instant explanations, solve problems, and generate new practice questions, effectively simulating on-demand tutoring support \cite{Kasneci2023ChatGPTFG, baidoo2023education}. While these capabilities present significant opportunities for supporting student learning, they also raise important pedagogical concerns. In particular, unstructured use of AI may encourage passive consumption of solutions rather than active engagement with underlying concepts, potentially limiting deep learning \cite{kirschner2006minimal, chi2009active}.

In physics education, where conceptual understanding and problem-solving skills are tightly intertwined, the manner in which students interact with AI tools is especially critical. Prior work in physics education research has consistently emphasized the importance of active learning, conceptual engagement, and deliberate practice in developing expertise \cite{hake1998interactive, freeman2014active}. Simply providing answers—even detailed ones—does not necessarily promote these processes. Therefore, there is a need for structured approaches that guide students to use AI as a tool for inquiry, practice, and conceptual development rather than as a shortcut to solutions.

To address this challenge, we introduce the Physics AI-Replication (P-A.I.R.) framework, a structured pedagogical approach designed to transform AI use into an active learning cycle rather than an answer-generating tool. The central mechanism of P-A.I.R. is problem replication: after prompting an AI system to explain a challenging concept and solve a target problem, the student requests new, similar practice problems, attempts them independently, and only then compares their solutions with those generated by the AI. This iterative process—identify, explain, replicate, reflect—is designed to promote retrieval practice and elaborative engagement, both of which have been consistently linked to durable conceptual understanding in physics education research~\cite{hake1998interactive, freeman2014active}.

The present study examines how students perceive the impact of the P-A.I.R. framework in introductory physics courses. Specifically, we analyze student-reported experiences related to conceptual understanding, problem-solving, efficiency, and confidence, as well as their willingness to continue using AI-assisted approaches. By focusing on student perceptions rather than direct performance measures, this work provides an initial evaluation of the feasibility and perceived educational value of structured AI integration in physics instruction. Given the increasing presence of AI tools in academic settings, understanding how to incorporate them in ways that support active learning while addressing challenges such as trust and sustained adoption is both timely and necessary.

%========================================================================================================================================================%
\section{Materials and Methods}
\label{method}
%========================================================================================================================================================%
This study was conducted at SUNY Farmingdale State College, a medium-sized public four-year institution in Farmingdale, New York.  Data were collected from three sections of algebra-based introductory physics, comprising two sections of Physics I and one section of Physics II. Each section enrolled 24 students ($N_{\text{enrolled}} = 72$). Participation in the research survey was voluntary and independent of course grading. 

All research procedures were reviewed and approved by the institutional review board at the author's institution, ensuring compliance with ethical guidelines for research involving human participants. Informed consent was obtained from all participants, who were assured of the confidentiality and anonymity of their responses. Data were collected and stored securely, and participants were informed of their right to withdraw from the study at any time without penalty.

The Physics AI-Replication (P-A.I.R.) framework was introduced at the beginning of the semester, and students were subsequently encouraged --- though not required --- to apply it throughout the course when encountering challenging homework or practice quiz problems.  A structured prompt was provided to all students to standardize the interaction. The protocol proceeded in four steps: (1)~the student identified a challenging problem from course materials; (2)~prompted an AI tool (of their choice, e.g., ChatGPT, Gemini) to explain the underlying concept and solve the problem step-by-step; (3)~requested two new, similar practice problems from the AI; and (4)~attempted those problems independently before reviewing the AI-generated solutions. This process was designed to promote active engagement through practice and reflection.  Each use of the protocol was accompanied by a brief optional reflection entry submitted via the course learning management system. Students were encouraged to describe what they learned, any remaining difficulties, and how the AI-generated practice problems influenced their understanding. These reflections were intended to support metacognitive engagement and instructional implementation but were not analyzed in the present study.  Students were permitted to submit up to one reflection per instructional unit (e.g., chapter), providing a structured but flexible opportunity to engage with the P-A.I.R. process throughout the semester. Students received a small amount of extra credit for completing reflections (up to 3\% of the final course grade), independent of research participation. Completion of the research survey, however, was entirely voluntary and carried no academic incentive.

The survey was administered via Qualtrics near the end of the semester, after students had sufficient opportunity to engage with the P-A.I.R. process. The instrument comprised three sections. Section~A collected background information, including AI usage frequency, primary purposes for using AI in physics, self-reported confidence in using AI effectively, expected GPA range, intended major, and AI tools used. Section~B contained seven experience items assessed on ordinal Likert scales: (B1)~perceived conceptual understanding (4-point scale), (B2)~helpfulness of replicated practice problems (4-point), (B3)~ease of AI interaction (5-point), (B4)~likelihood of continued use (5-point), (B5)~perceived quality and accuracy of AI outputs (4-point), (B6)~time efficiency relative to traditional study methods (5-point), and (B7)~likelihood of recommending the P-A.I.R.\ approach to peers (5-point). Section~C included one numerical confidence item (C1: self-rated problem-solving confidence on a 1--10 scale) and two open-ended questions: (C2)~one specific way P-A.I.R.\ changed their approach to studying physics, and (C3)~one suggested improvement to the process. The survey was anonymous, and IP address collection was disabled in Qualtrics. The investigator did not access survey data until after final grades were submitted.

Quantitative survey responses were analyzed using descriptive statistics implemented in Python to examine overall trends in student perceptions across survey items. Given the ordinal nature of Likert-scale data, non-parametric statistical tests were employed where appropriate. Qualitative responses to open-ended questions were analyzed using an inductive thematic approach. Responses were reviewed to identify recurring patterns and themes, following established guidelines for thematic analysis~\cite{braun2006using}. All responses were coded by the author, and representative excerpts are reported with minor typographical corrections where necessary.

%========================================================================================================================================================%
\section{Results}
\label{results}
%========================================================================================================================================================%
Of the 72 enrolled students, 39 returned surveys (response rate: 54.2\%), though not all respondents answered every question. Participants were predominantly from engineering and engineering  technology disciplines — including Architectural Engineering, Civil Engineering, Computer Science, Cybersecurity, and Mechanical Engineering — representing approximately 80\% of the sample  ($n = 31$). The remaining participants came from Bioscience; Science, Technology \& Society; or Liberal Arts backgrounds ($n = 8$). The sample included 10 female students, with the remainder identifying as male. Prior to the intervention, students reported frequent use of AI tools for academic purposes. Over half of respondents (51.3\%) reported using AI tools often (2–3 times per week), and an additional 15.4\% reported daily use; only one participant (2.6\%) indicated rare use. ChatGPT was the most commonly used platform (82.0\%), followed by Google Gemini (41.0\%), with many students reporting use of more than one AI tool. Students reported using AI for a range of academic tasks, most commonly for concept explanation (84.6\%) and problem-solving assistance (79.5\%), followed by exam preparation (71.8\%) and homework support (64.1\%). Despite frequent use, self-rated confidence in using AI \textit{effectively} for physics was largely moderate: 48.7\% reported moderate confidence, 33.3\% high confidence, and only 10.3\% extreme confidence.

\begin{figure}[htb!]
    \centering
    \includegraphics[width=0.9\columnwidth]{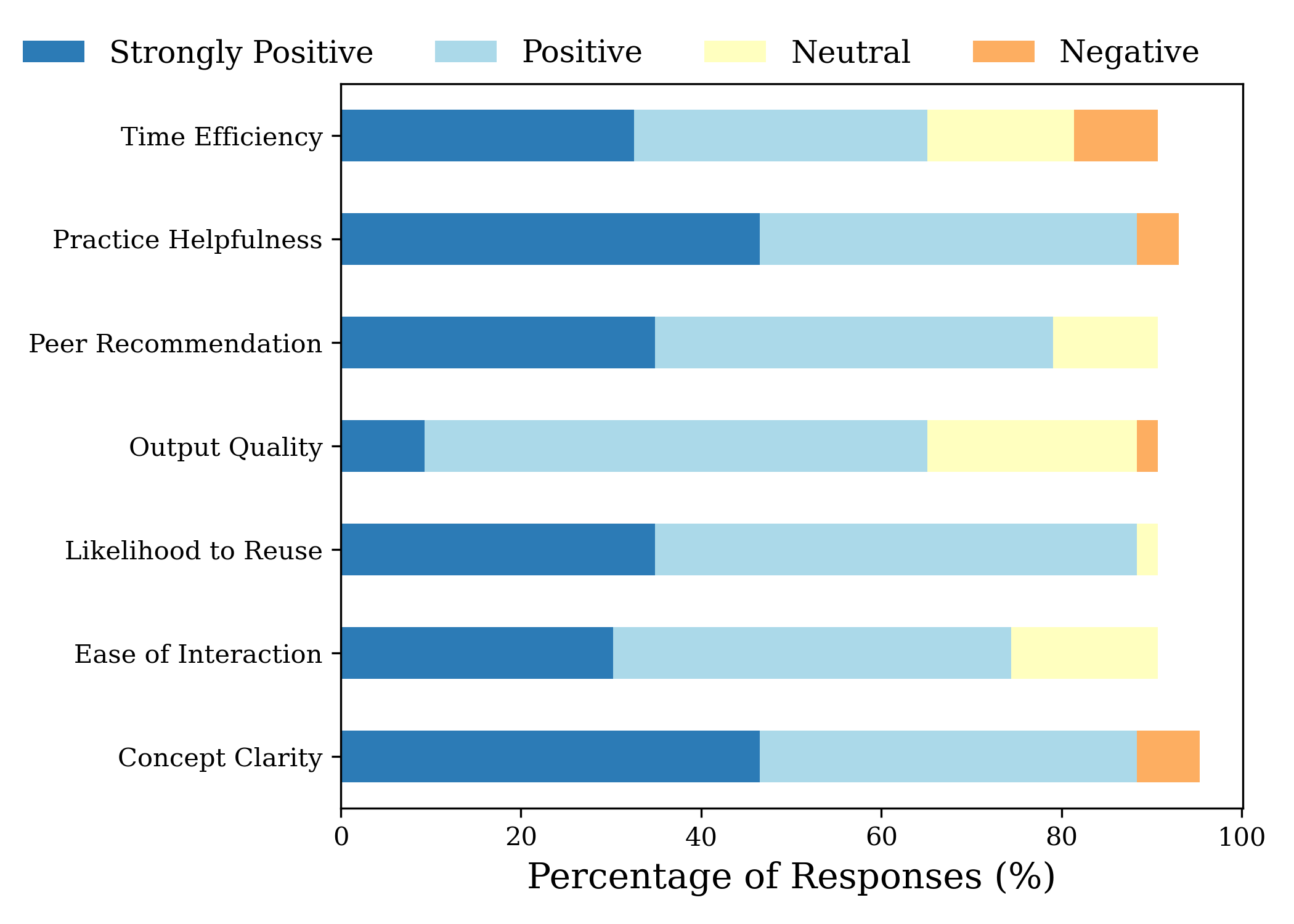}
    \caption{Distribution of student responses to Section B experience items (B1--B7). Responses are strongly skewed toward positive categories across all items, with particularly high ratings for conceptual understanding (B1) and problem-replication helpfulness (B2).}
    \label{fig:likert_summary}
\end{figure}

Responses to the P-A.I.R. experience items were strongly positive overall (Table~\ref{tab:likert}). As shown in Fig.~\ref{fig:likert_summary}, responses to the conceptual understanding item (B1) were highly skewed toward positive categories, with nearly all participants reporting improvement: 51.3\% rated their understanding as \textit{significantly improved} and 46.2\% as \textit{somewhat improved} ($\mu = 2.49$, $\sigma = 0.56$). A similarly strong pattern was observed for the problem-replication component (B2), with 51.3\% rating it \textit{very helpful} and 46.2\% \textit{somewhat helpful}, indicating that both explanation and practice elements were perceived as beneficial. A Spearman rank correlation confirmed a strong association between conceptual clarity and replication helpfulness ($r_s = 0.582$, $p < 0.001$, $n = 38$), suggesting that students who found AI explanations clearer also benefited more from solving replicated problems. This relationship is consistent with the intended sequential structure of the P-A.I.R. framework, in which conceptual understanding facilitates productive practice. 

Across the remaining experience items (B3--B7), responses indicate generally positive but more variable perceptions of practical usability and implementation. Students found the AI interaction accessible: 69.2\% rated ease of use as \textit{somewhat} or \textit{extremely easy} (B3). Output quality (B5) was rated \textit{good} or \textit{excellent} by 71.8\%, while 25.6\% rated it \textit{fair}, reflecting a pragmatic recognition of AI limitations. Relative to traditional study methods, 68.6\% of respondents found the AI approach more time-efficient (B6), with 40.0\% rating it \textit{much more efficient}. Adoption-related measures were also high: 89.7\% expressed likelihood of continued use (B4), and 82.1\% would recommend the P-A.I.R.\ framework to peers (B7). A significant positive association was observed between likelihood of reuse and peer recommendation ($r_s = 0.42$, $p = .007$, $n = 39$), indicating that students who intended to use the framework again were also more likely to recommend it to others. Together, these findings indicate that students perceived strong conceptual benefits and expressed both personal and social endorsement of the P-A.I.R. framework.

\begin{table}[htb!]
\caption{\label{tab:likert}%
Summary statistics for Section B experience items. \% pos.\ = percentage of valid responses at or above the positive threshold. Items were measured on varying Likert-type scales and coded numerically for descriptive analysis.}
\begin{ruledtabular}
\begin{tabular}{lcccc}
Item & $n$ & $\mu$ & $\sigma$ & \% pos. \\
\hline
B1 Concept clarity         & 39 & 2.49 & 0.56 & 97.4 \\
B2 Practice helpfulness    & 38 & 1.53 & 0.51 & 97.4 \\
B3 Ease of interaction     & 39 & 2.90 & 1.02 & 69.2 \\
B4 Likelihood to reuse     & 39 & 4.15 & 0.90 & 89.7 \\
B5 Output quality          & 39 & 2.79 & 0.66 & 71.8 \\
B6 Time efficiency         & 35 & 2.97 & 1.04 & 68.6 \\
B7 Peer recommendation     & 39 & 3.15 & 0.84 & 82.1 \\
%C1 Self-confidence (1--10) & 32 & 7.03 & 1.18 & ---  \\
\end{tabular}
\end{ruledtabular}
\end{table}

\begin{figure}[h]
\includegraphics[width=0.85\columnwidth]{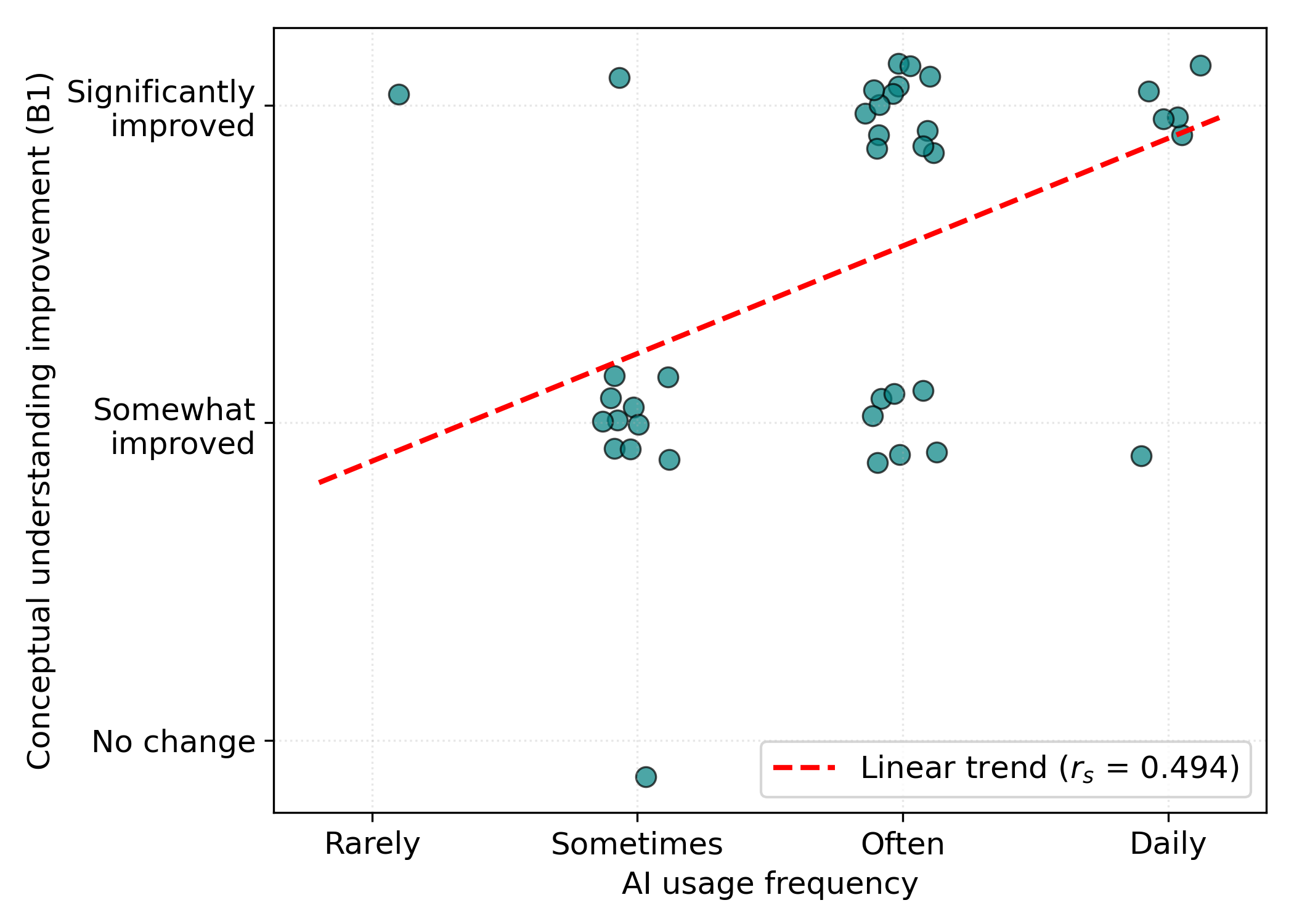}
\caption{\label{fig:freq_clarity} AI usage frequency vs.\ concept  clarity improvement (B1). Each point represents an individual student response. The dashed line indicates the linear trend ($r_s = 0.494$, $p = 0.001$, $n = 39$.)}
\end{figure}

AI usage frequency was significantly associated with perceived concept clarity ($r_s = 0.494$, $p = 0.001$, $n = 39$;  Fig.~\ref{fig:freq_clarity}), suggesting that students who used AI tools more frequently reported greater perceived improvements in conceptual understanding. Because these data are self-reported, this association should be interpreted with caution.

\begin{figure}[htb!]
    \centering
    \includegraphics[width=0.85\columnwidth]{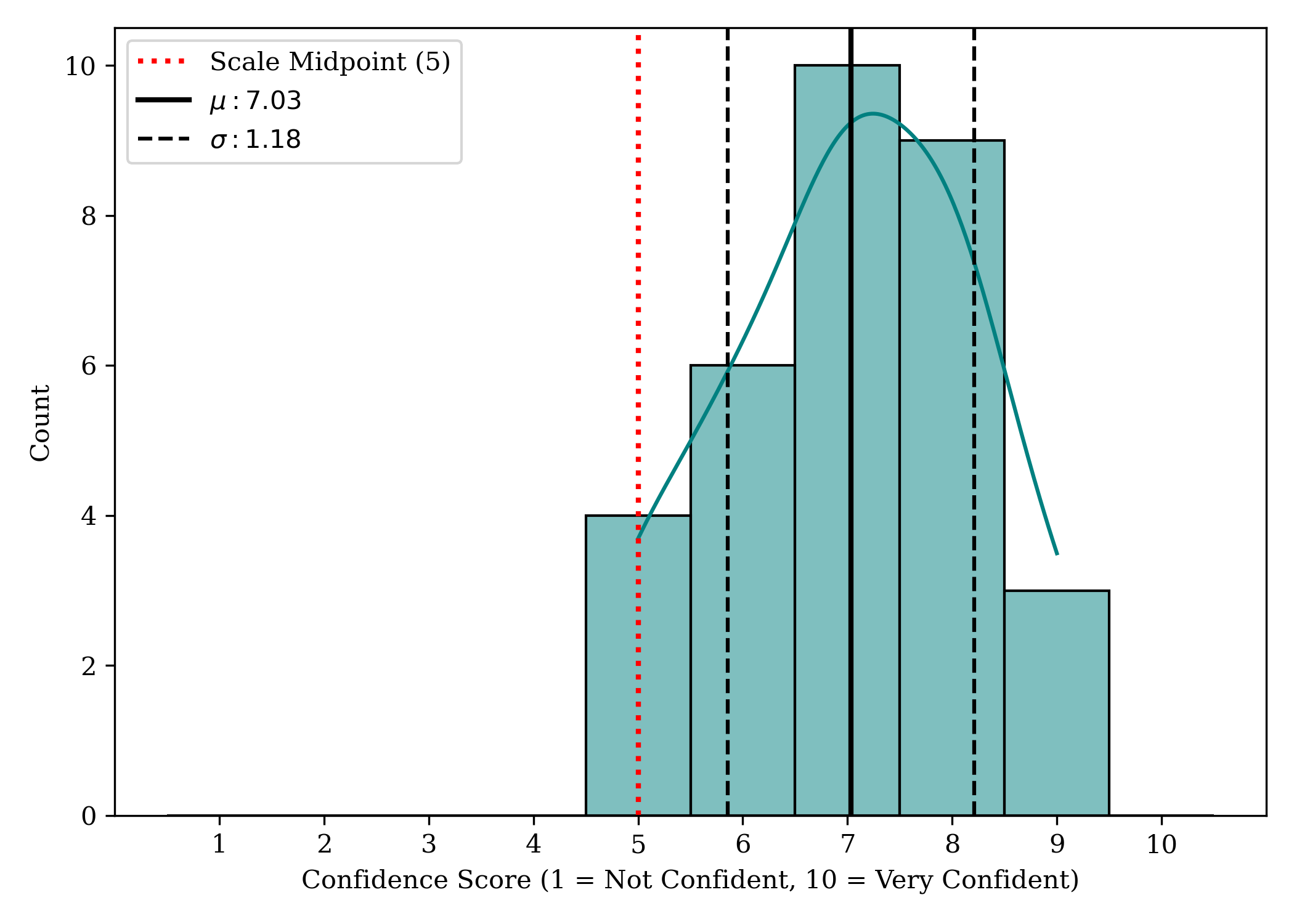}
    \caption{Distribution of post-engagement self-confidence scores (C1; 1–10 scale). Scores are concentrated above the midpoint of the scale (dashed line at 5), with no responses below 5, indicating generally high perceived ability to solve similar problems independently after engaging with the P-A.I.R. framework.}
    \label{fig:confidence_dist}
\end{figure}

\begin{figure}[htb!]
\centering
\includegraphics[width=0.85\columnwidth]{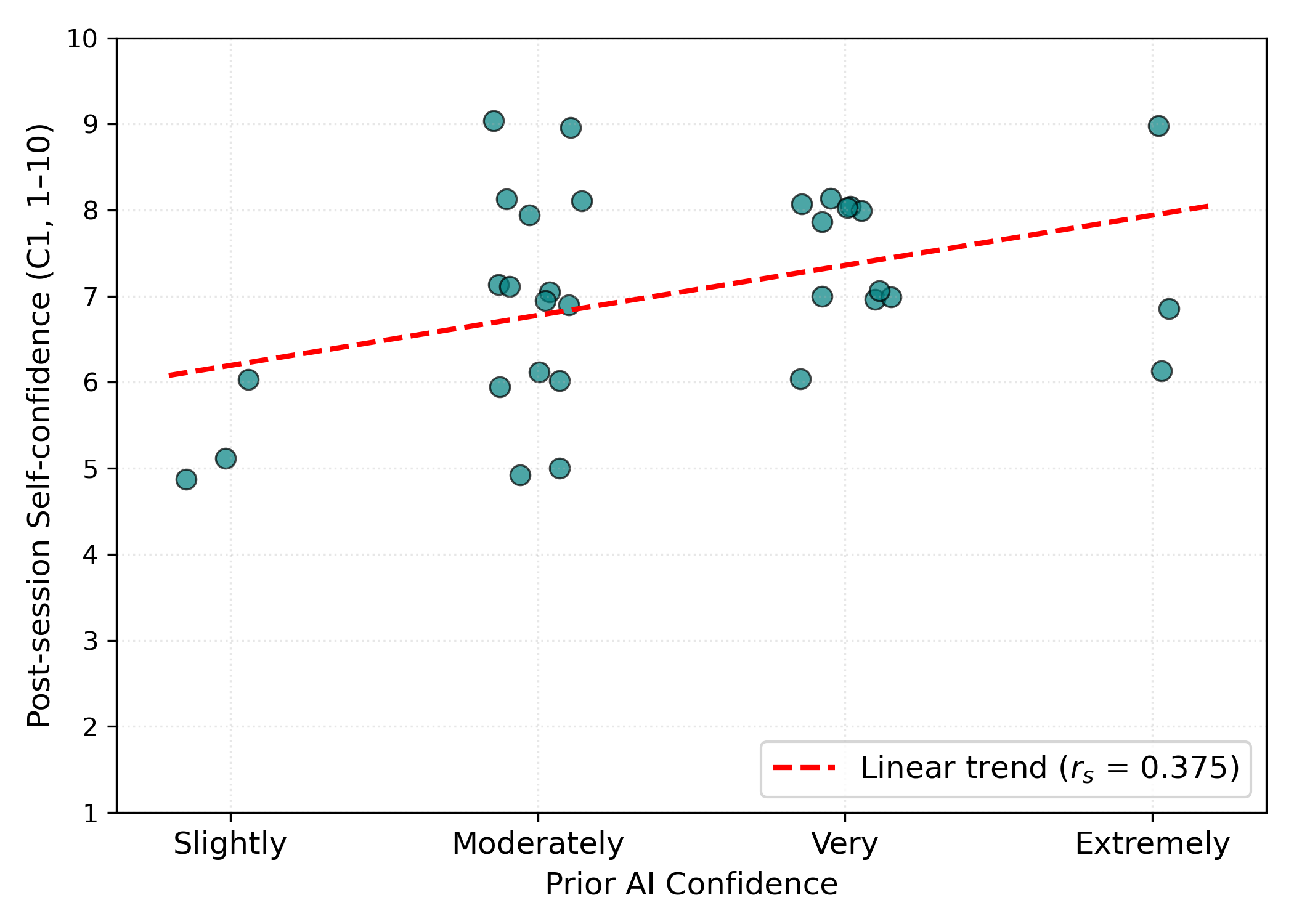}
\caption{Prior AI confidence vs.\  self-confidence score (C1). Each point represents an individual student response. The dashed line indicates the linear trend ($r_s = 0.375$, $p = 0.034$, $n = 32$.)}
\label{fig:conf_selfconf}
\end{figure}

Self-confidence (C1; 1--10 scale) was reported by 32 participants ($\mu = 7.03$, $\sigma = 1.18$, $\text{Mdn} = 7.0$; range 5--9). As shown in Fig.~\ref{fig:confidence_dist}, confidence scores were consistently above the midpoint of the scale, with no responses below 5, indicating generally high self-reported confidence in solving similar problems independently after engaging with the P-A.I.R. framework. Additionally, prior AI confidence was significantly associated with  post-session self-confidence scores ($r_s = 0.375$, $p = .034$,  $n = 32$; Fig.~\ref{fig:conf_selfconf}), suggesting that students who felt more capable using AI tools also reported higher confidence in their ability to solve physics problems independently after engaging with the P-A.I.R.\ framework. Together, these results indicate that both prior familiarity with AI and confidence in its use are linked to more favorable learning-related perceptions following engagement with P-A.I.R.

Qualitative responses ($n = 34$ for C2; $n = 33$ for C3)  revealed several consistent themes regarding how the P-A.I.R. framework influenced students’ learning. The most prominent theme was \textit{conceptual clarification}, with several students describing how AI explanations simplified complex ideas and made physics concepts more accessible. Students frequently noted that AI helped translate formal or abstract material into more intuitive explanations, often in a conversational or step-by-step format,  functioning similarly to an instructor or tutor providing immediate, simplified guidance.

A second major theme was \textit{problem decomposition and structured reasoning}. Students reported that the P-A.I.R. process helped them break problems into manageable steps and better understand how to apply equations in context. This aligns with the quantitative finding that conceptual clarity was strongly associated with the perceived helpfulness of replicated practice. A notable metacognitive  dimension also emerged, with several students describing how crafting  effective AI prompts required them to first understand the problem structure themselves --- reflecting active engagement rather than passive consumption of AI outputs. A related benefit was \textit{autonomous practice generation}, with students valuing the ability to produce unlimited, customized problems on demand rather than relying solely on course materials. Together, these themes also contributed to \textit{increased efficiency}, as students reported working through problems more quickly while maintaining understanding.

At the same time, responses to the improvement question (C3) revealed important limitations. Students identified two principal concerns: first, a mismatch between AI-generated solutions and in-class methods, with AI sometimes introducing notation or approaches not covered in the course; and second, accuracy and trust, with students noting that AI-generated solutions occasionally contained errors and required verification. Notably, one student reported no change in studying approach, serving as a reminder that perceived benefits were not universal. Many students suggested incorporating human oversight or instructor validation to improve reliability, with additional suggestions including visual explanations, greater problem variation, and more sophisticated AI tools.

Overall, the qualitative findings support the quantitative results, indicating that students perceived the P-A.I.R. framework as helpful for conceptual understanding and problem-solving, while also highlighting the importance of accuracy and instructional alignment for effective implementation.

%========================================================================================================================================================%
\section{Discussion and Conclusions}
\label{disc}
%========================================================================================================================================================%
The results provide converging evidence that students perceived the P-A.I.R. framework as conceptually beneficial and practically useful, with strong willingness to adopt it in future learning. Critically, nearly all participants (97.4\%) reported some degree of perceived conceptual improvement after engaging with the P-A.I.R.\ framework, and  self-confidence scores clustered well above the scale midpoint ($\mu = 7.03$; range 5--9), with no student reporting a score below 5 --- supporting the feasibility of P-A.I.R.\ as a low-barrier supplement to traditional instruction.

The strong association between conceptual clarity and replication helpfulness ($r_s = 0.582$, $p < 0.001$) supports the intended sequential structure of P-A.I.R., consistent with cognitive load theory~\cite{kirschner2006minimal}, which predicts that worked examples free cognitive resources for schema formation. The qualitative theme of metacognitive engagement --- students describing the need to understand problem structure before constructing effective AI 
prompts --- further suggests that P-A.I.R.\ promotes deeper processing beyond passive AI consultation, consistent with prior research on active and constructive learning processes~\cite{chi2009active, hake1998interactive, freeman2014active}.

Prior experience with AI also appears to play a role. More frequent AI use was associated with greater perceived conceptual benefit, while confidence in using AI was associated with higher post-engagement self-confidence. However, because these measures are self-reported, the observed associations may partially reflect students' pre-existing attitudes toward AI rather than differences in their learning experiences. Students also reported that P-A.I.R. was generally efficient and usable, with high likelihood of reuse and recommendation. The relationship between these measures indicates that positive experiences translated into both personal adoption and peer endorsement, supporting the feasibility of integrating structured AI use into regular study practices.

At the same time, qualitative responses revealed two actionable limitations. First, students noted a mismatch between AI-generated solutions and in-class methods, suggesting that instructor-designed prompt scaffolds could improve alignment with course expectations. Second, concerns about AI accuracy, consistent with prior findings~\cite{Paul2026APAI}, highlight the importance of verification steps and instructor oversight in future implementations. Together, these findings indicate that while students value AI for explanation and guidance, they remain cautious about relying on it independently.

We acknowledge several limitations of this work. The study relies on self-reported perceptions rather than direct learning outcomes, and the absence of a pre-intervention measure means that perceived gains cannot be attributed to the P-A.I.R. framework with certainty. The sample was drawn from a single institution and was skewed toward engineering majors, which may limit the generalizability of the findings. Participation in the survey was voluntary, and respondents who elected to complete it may have been more engaged with the P-A.I.R. framework—or with coursework generally—than non-respondents, potentially limiting the representativeness of the findings. Future work should examine objective learning outcomes and explore how instructional supports influence the effectiveness of structured AI use.

Overall, the findings suggest that P-A.I.R.\ can transform a tool students already use frequently --- but often inconsistently --- into a structured framework that supports active learning, guided practice, and independent problem-solving. As AI becomes increasingly embedded in student study habits, frameworks that guide rather than restrict its use offer a pragmatic and pedagogically grounded path forward for physics instruction.

%\section*{Conflict of Interest}
%The authors declare that the research was conducted without any commercial or financial relationships that could be construed as a potential conflict of interest.
\acknowledgments{}
%The authors extend their gratitude to their students for their participation in this study. 
T%he author gratefully acknowledges Prof. Xu Zhang of the Department of Economics, SUNY Farmingdale State College, for helpful discussions. 
The author also extends sincere gratitude to the students who participated in this study.

\bibliographystyle{unsrt}
\bibliography{main}

\end{document}